\documentclass[twocolumn,showpacs,amsmath,amssymb,aps,prl,superscriptaddress]{revtex4-1}

\usepackage[]{cleveref}
\usepackage[]{graphicx}
\usepackage[]{verbatim}
\usepackage[]{color}
\usepackage[abs]{overpic}
\usepackage{rotating}
\usepackage{tabularx}

\usepackage{amsfonts}
\usepackage{amsmath}
\usepackage{amssymb}
\usepackage{bm}
\usepackage{graphicx}
\usepackage[breaklinks,colorlinks=true]{hyperref}
\usepackage{mathrsfs}
\usepackage[lofdepth,lotdepth,caption=false]{subfig}
\usepackage{varwidth}
\usepackage{wrapfig}
\usepackage{times}
\usepackage{longtable}
\usepackage{multirow}
\usepackage{tikz}
\definecolor{darkblue}{RGB}{0 60 120}
\definecolor{eggplant}{RGB}{190 10 150}
\definecolor{darkgray}{RGB}{70 70 70}
\definecolor{lightgray}{RGB}{80 80 80}
\definecolor{lightgray2}{RGB}{245 215 110}
\definecolor{lightgray3}{RGB}{255 0 0}

\newcommand{\blio}{$\beta$LIO}
\newcommand{\aaio}[1]{$\alpha$#1IO}
\newcommand{\io}[2]{$#1$-#2$_2$IrO$_3$}

\newcommand{\jeff}{$j_{\text{eff}}=1/2$}

\graphicspath{{./}}

\begin{document}

\title{Predominance of the Kitaev interaction in a three-dimensional honeycomb iridate: \\
from \textit{ab-initio} to spin model}

\author{Heung-Sik Kim}
\affiliation{Department of Physics and Center for Quantum Materials , University of Toronto, 60 St.~George St., Toronto, Ontario, M5S 1A7, Canada}

\author{Eric Kin-Ho Lee}
\affiliation{Department of Physics and Center for Quantum Materials , University of Toronto, 60 St.~George St., Toronto, Ontario, M5S 1A7, Canada}

\author{Yong Baek Kim}
\affiliation{Department of Physics and Center for Quantum Materials , University of Toronto, 60 St.~George St., Toronto, Ontario, M5S 1A7, Canada}
\affiliation{Canadian Institute for Advanced Research / Quantum Materials Program, Toronto, Ontario MSG 1Z8, Canada}

\begin{abstract}
  The recently discovered three-dimensional hyperhoneycomb iridate,
  \io{\beta}{Li}, has raised hopes for the realization of dominant
  Kitaev interaction between spin-orbit entangled local moments due to
  its near-ideal lattice structure. If true, this material may lie
  close to the sought-after quantum spin liquid phase in three
  dimensions. Utilizing \textit{ab-initio} electronic structure
  calculations, we first show that the spin-orbit entangled basis,
  \jeff{}, correctly captures the low energy electronic structure. The
  effective spin model derived in the strong coupling limit
  supplemented by the \textit{ab-initio} results is shown to be
  dominated by the Kitaev interaction. We demonstrated that the
  possible range of parameters is consistent with a non-coplanar
  spiral magnetic order found in a recent experiment. All of these
  analyses suggest that \io{\beta}{Li} may be the closest among known
  materials to the Kitaev spin liquid regime.
\end{abstract}

\maketitle

\textit{Introduction --}
Kitaev's exact solution of a quantum
spin-liquid on a spin-$1/2$ honeycomb model has spurred considerable
interest in the search for a material
realization\cite{kitaev2006anyons,singh2012relevance}.  Of particular
focus is the family of quasi-two-dimensional (2D) honeycomb iridate
materials \io{\alpha}{$A$} ($A=\text{Na, Li}$, hereafter \aaio{$A$}),
where iridium (Ir) ions form decoupled layers of honeycomb
lattices\cite{choi2012spin,Ye2012dr} and have been argued to host
spin-orbital entangled \jeff{} degrees of
freedom\cite{kim2008novel,shitade2009quantum,Gretarsson2013fp}. Due to
the interplay of strong atomic spin-orbit coupling (SOC) and
correlation effects, these \jeff{} moments in the ideal \aaio{$A$}
structure interact in the highly anisotropic manner described by the
Kitaev model\cite{jackeli2009mott}.  In addition to these Kitaev-type
exchanges, the symmetries of the ideal structure also permit
additional exchanges that generate a plethora of interesting phases of
matter\cite{rau2014generic}. In reality, however, these materials
possess sizeable monoclinic distortions that deform the octahedral
oxygen cages surrounding Ir ions\cite{Ye2012dr,OMalley2008p435}.
These distortions lower the symmetry of the system and therefore
complicate the description of these materials.  Thus far, a consensus
on the minimal model required to describe this family of 2D honeycomb
iridates has yet been reached; a distortion-free analog of these
honeycomb iridates may offer a more direct path towards the
realization of Kitaev physics.

\begin{figure}
  \centering
  \includegraphics[width=0.45 \textwidth]{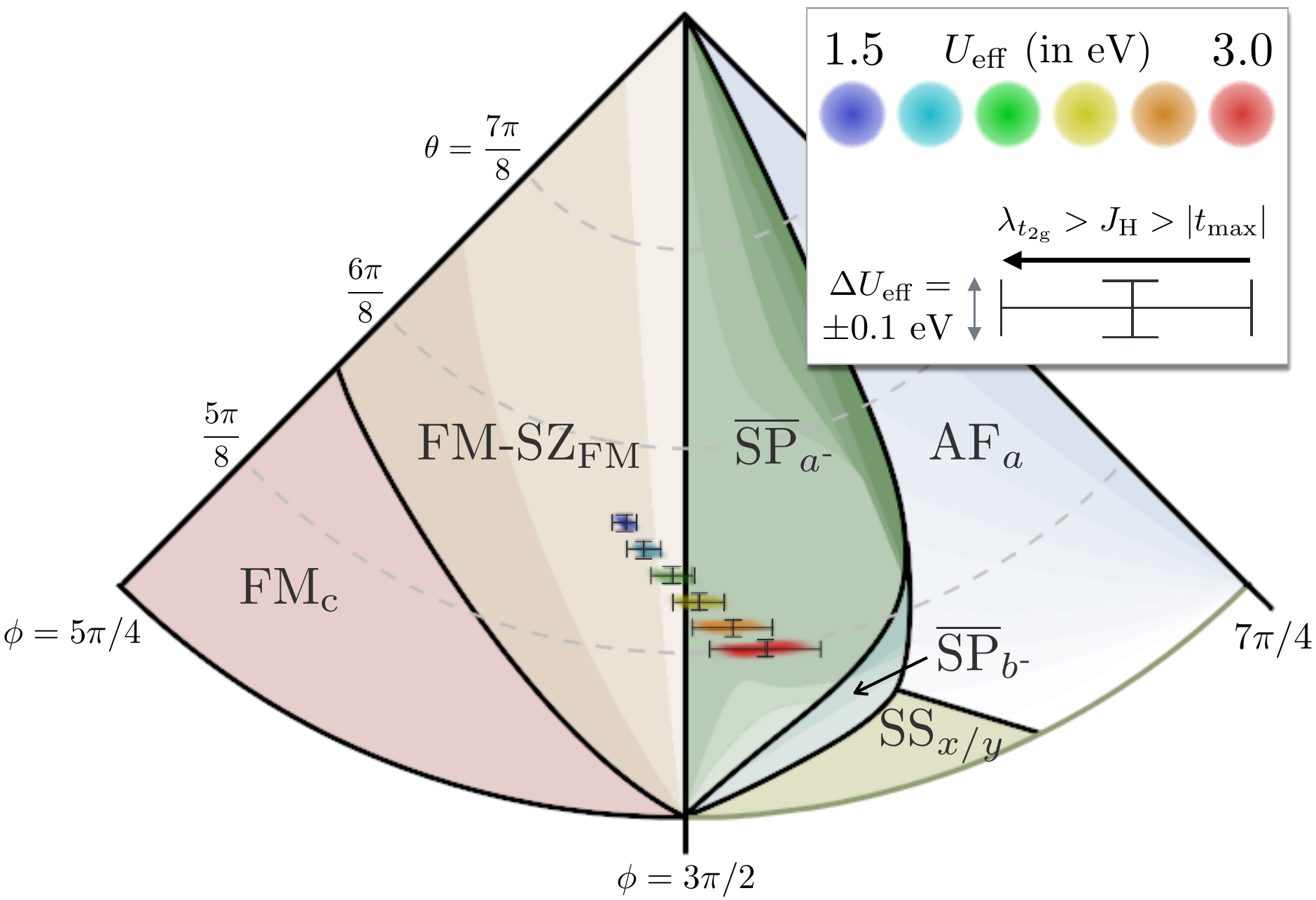}
  \caption{(Color online) Phase diagram of the $J$-$K$-$\Gamma$ model
    reproduced from Ref. \cite{Eric2014tm}, overlaid with density
    distributions of exchange interaction parameters estimated from
    \textit{ab-initio} results for \blio{}; see main text for details.
    Six shaded areas with different colors represent relevant $U_{\rm eff}$
    values ranging from 1.5 to 3.0 eV in increments of 0.3 eV. 
    Vertical and horizontal markers on the shaded areas
    depict the variation of $U_{\rm eff}$ and $J_{\rm H}$, respectively,
    as shown in the inset.
    Note that, when $U_{\rm eff} \geq 2.4$ eV, corresponding phases 
    lie within the green $\overline{\text{SP}}_{a^{\textrm{-}}}$
    spiral phase area. $\overline{\text{SP}}_{a^{\textrm{-}}}$ phase 
    is consistent with the magnetic order observed in the
    experimental work of Ref. \cite{Biffin2014ky}. For detailed
    discussion of the other phases, see Ref. \cite{Eric2014tm}.}
  \label{fig:magldau}
\end{figure}

The timely discovery and synthesis\cite{Biffin2014ky,Takayama2014uf}
of the hyperhoneycomb \io{\beta}{Li} (hereafter \blio{}) may present
such an exciting opportunity. Much like its 2D counterpart, the Kitaev
model on the ideal, 3D hyperhoneycomb lattice supports an exact spin
liquid ground
state\cite{Mandal2009es,Eric2014hk,kimch2013td,Nasu2014ft}.  In
addition, the distortion-free, classical pseudospin-$1/2$ model on the
hyperhoneycomb lattice also supports a myriad of complex magnetic
phases\cite{Eric2014tm}. Moreover, interesting topological phases have
been predicted on this lattice\cite{Schaffer2014ts}.  These previous
results illustrate the possibilities that may be realizable in
\blio{}; however, they rely on the use of the \jeff{} degrees
of freedom in the low-energy description of \blio{}, which has not
been justified microscopically.  Furthermore, whether the near-ideal
structure of \blio{} can give rise to a simple minimal pseudospin
model dominated by the Kitaev exchange has so far not been validated.
Also, with the recent experimental observation of a magnetic spiral
order in \blio{}, any minimal model and its accompanying parameters
must also be capable of predicting the observed order: this provides a
stringent test of feasibility for any model describing \blio{}.

\begin{figure}
  \centering
  \includegraphics[width=0.40 \textwidth]{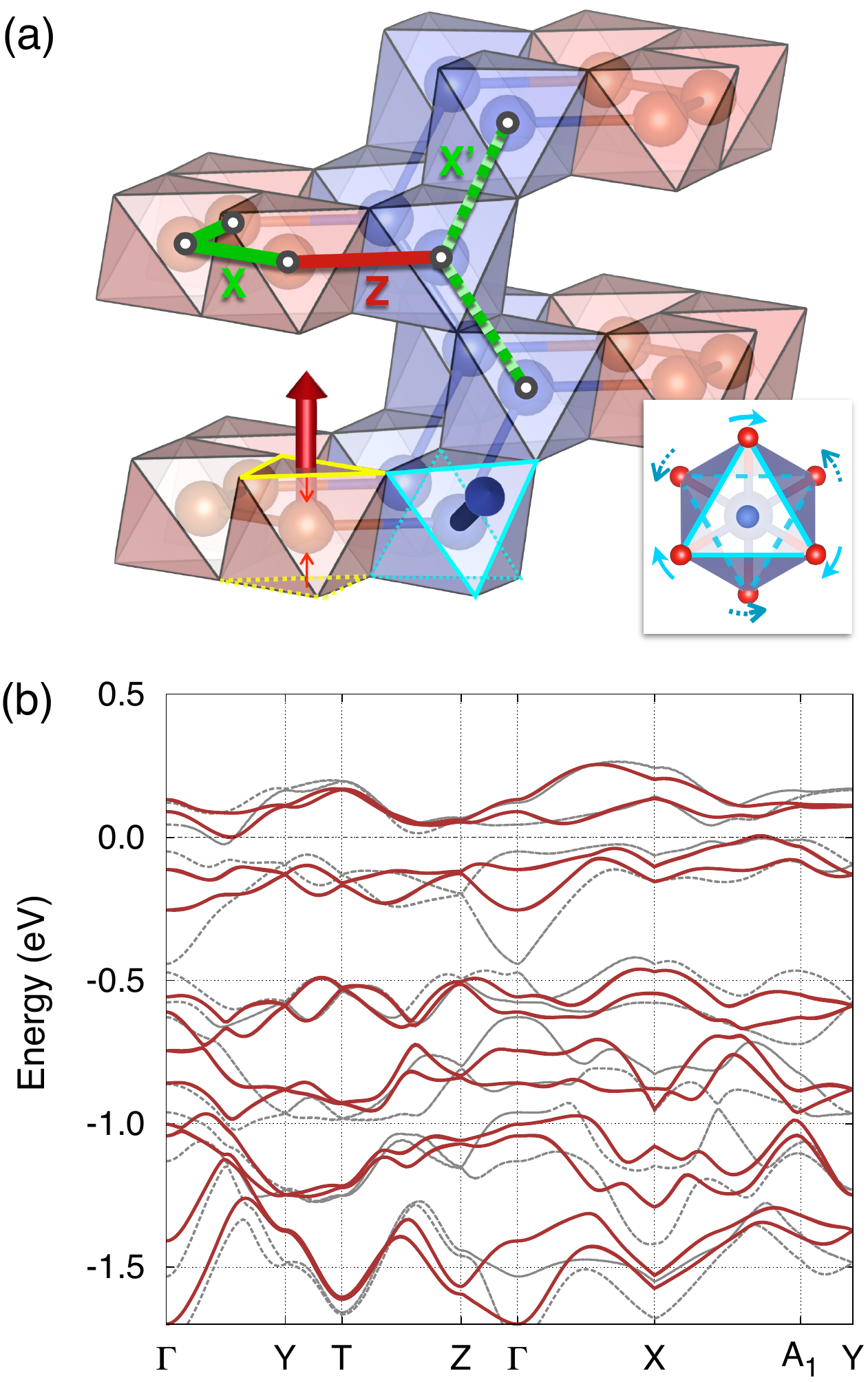}
  \caption{(Color online) (a) Network of IrO$_6$ octahedra in the
    hyperhoneycomb lattice.  The two distinct nearest-neighbor (NN)
    bonds, X and Z, are depicted as solid green and red lines,
    respectively.  X and X' bonds are symmetry equivalent, whereas Z
    bonds are distinct.  Emerging from the octahedra, the red and blue
    arrows point in the direction of the trigonal distortions for red
    and blue IrO$_6$ octahedra respectively.  The trigonal distortion,
    which consists of the compression and rotations of the opposing
    oxygen trangles, is illustrated in the figure and in the inset.
    (b) shows the band structure with
    the presence of spin-orbit coupling (SOC).  Solid red and dashed
    grey curves are the band structure of the experimental and ideal
    structures, respectively.}
  \label{fig:struct}
\end{figure}

\begin{figure*}
  \centering
  \includegraphics[width=0.98 \textwidth]{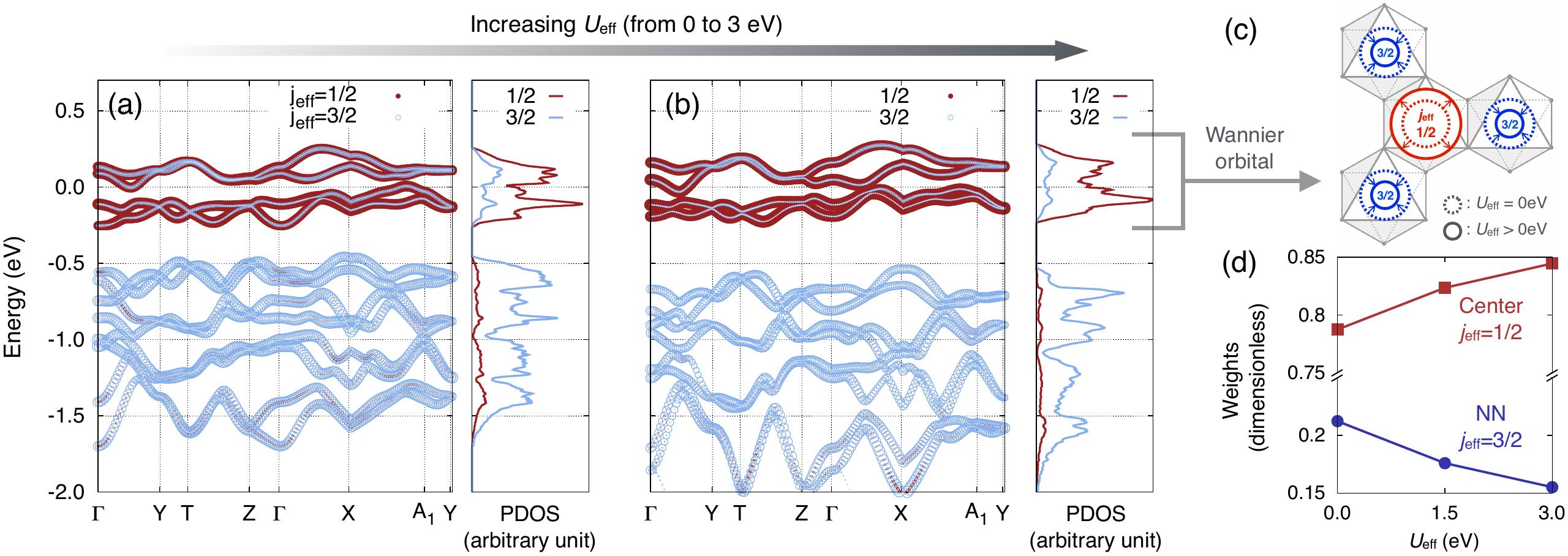}
  \caption{(Color online) (a,b) Band structure and density of states
    (DOS) projected onto the $j_{\rm eff}$ states in the presence of
    SOC (a) without and (b) with the on-site Coulomb interaction
    $U_{\rm eff}$ = 3.0~eV. (c) shows the schematic shape of the
    $j_{\rm eff}$=1/2-like Wannier orbital constructed from the
    $j_{\rm eff}$=1/2-dominated bands near the Fermi level.  Dashed
    and solid circles depict the Wannier orbitals from calculations
    without and with finite $U_{\rm eff}$, respectively.  Weights of
    the central $j_{\rm eff}$=1/2 and nearest-neighboring $j_{\rm
      eff}$=3/2 tail in the orbital are shown in (d) as a function of
    $U_{\rm eff}$.  }
  \label{fig:jbands}
\end{figure*}

In this letter, we tackle these issues by combining results of our
\textit{ab-initio} electronic structure calculations and a
strong-coupling theory to arrive at a \jeff{} model to describe
\blio{}.  From our \textit{ab-initio} band structure results, we find
that the low-energy states can be described in terms of localized
\jeff{} states because of the large atomic SOC present in Ir.  In
fact, the magnitude of SOC in the paramagnetic state is enhanced by
the electron interactions in Ir $d$ orbitals, which is consistent with
recent observations in several 4$d$ and 5$d$ transition metal
compounds\cite{Liu2008SRO,hck_arxiv,Vijay2014}. To go beyond the
limitation of {\it ab-initio} calculations in treating electron
interactions, we employ the strong-coupling expansion recently
proposed in \cite{rau2014generic} to arrive at a minimal
pseudospin-$1/2$ model.  Due to the near-ideal structure of \blio{},
we discover that the resulting pseudospin model is near-isotropic
while both distortion-induced and further neighbor interactions are
small.  Remarkably, the estimated exchange interactions places the
model near the ferromagnetic Kitaev limit and within a region where
the classical ground state agrees well with the
experimentally-observed spiral phase\cite{Biffin2014ky}.  Our results
are summarized in Fig. \ref{fig:magldau} and elaborated in the rest of
this work.

\textit{Structure and ab-initio calculations --}
\blio{} is a member
of the generic three-dimensional (3D) harmonic honeycomb iridate
series\cite{Modic2014ch}, which are structural variants of
2D-honeycomb iridates \aaio{$A$}. The hyperhoneycomb lattice is
composed of a tri-coordinated network of edge-shared IrO$_6$ octahedra
as shown in Fig. \ref{fig:struct}(a).  There are two types of
nearest-neighbor (NN) bonds in this network: we denote these bonds as
X and Z.  Despite being symmetry-inequivalent, these two bonds are
almost identical owing to their similar local crystal structures as
revealed by recent structural
analysis\cite{Biffin2014ky,Takayama2014uf}.

The crystal structure refinement also revealed nearly ideal IrO$_6$
octahedra compared to those from \aaio{L}: standard deviations of the
Ir-O bond length and O-Ir-O bond angles in \blio{} (0.002~\AA{} and
2.77$^{\circ}$) are much smaller than those in \aaio{L} (0.050~\AA{}
and 4.56$^{\circ}$)\cite{OMalley2008p435}.  Since finite standard
deviations in O-Ir-O bond angles is a result of trigonal distortion in
the IrO$_6$ octahedra, the small value present in \blio{} indicates
that trigonal distortions are indeed small in this compound (the
directions of trigonal distortion are shown as colored arrows in the
figure). The nearly ideal IrO$_6$ octahedra in \blio{} suggest that
local crystal fields are principally cubic in symmetry, therefore the
spin-orbital entangled \jeff{} states would be a good basis to
construct a low-energy description of this material in the presence of
strong SOC.

To validate the use of \jeff{} states in the low-energy description of
\blio{}, we turn to \textit{ab-initio} electronic structure
calculations\footnote{See Sec. A in \href{https://sites.google.com/site/heungsikim/home/temporary-files/SM_bLIO_resbmit_151023.pdf?attredirects=0&d=1}{Supplementary Materials} posted in 
the author's personal web page (\href{https://sites.google.com/site/heungsikim/home/temporary-files}
{https://sites.google.com/site/heungsikim/home/temporary-files}) for details.}. The band
dispersions of the ideal and experimental structures with SOC can be
seen in Fig. \ref{fig:struct}(b).  The dispersions from the
experimental structure (solid red curves) and those from ideal one
(dashed grey curves) share similar overall shape, especially near the
chemical potential.  The separation between the upper eight bands and
the lower sixteen bands (including Kramers degeneracies) can be
clearly seen in the figure, suggesting the formation of \jeff{} and
$j_{\text{eff}}=3/2$ bands\footnote{We comment that, due to the loss
  of chiral symmetry and the mixing of the $j_{\rm eff}=3/2$ states in
  the $j_{\rm eff}=1/2$ subspace, the nodal Fermi ring mentioned in
  previous tight-binding analysis is absent in both of the ideal and
  experimental structures}.

In Fig. \ref{fig:jbands}(a), we show the $j_{\rm eff}$-projected band
dispersions and density of state (PDOS) in the presence of SOC based
on the experimental structure. 
The Projection is done by taking inner products between 
the atomic $j_{\rm eff}$ states and the Bloch state represented in terms of
the local pseudo-atomic orbital basis. Weights of the $j_{\rm eff}$=1/2 and 3/2 
components within each Bloch state are depicted as the size of red and blue circles,
respectively, in the band plots.
The large \jeff{} PDOS weight in the
upper eight bands---the closest bands to the Fermi level---indicates
the development of \jeff{} bands and confirms that the basis states
relevant to the low-energy description of \blio{} possess mostly
\jeff{} character. The effect of electron correlations inherent to Ir
$t_{\rm 2g}$ orbital further enhances the \jeff{} character as shown
in Fig. \ref{fig:jbands}(b), where effective on-site Coulomb
interaction $U_{\text{eff}}\equiv U - J_H$ is included within the
DFT+$U$ formalism ($J_H$ is Hund's coupling; for details see Sec. A in
\href{https://sites.google.com/site/heungsikim/home/temporary-files/SM_bLIO_resbmit_151023.pdf?attredirects=0&d=1}{Supplementary Materials}).\footnote{In order to capture the role of
  electron correlations in the band structure only, we do not
  consider any magnetism in our DFT+$U$ calculations in this work.}
As $U_{\text{eff}}=3.0$~eV is added, the separation between the \jeff{}
and $j_{\rm eff}=3/2$ states becomes enlarged. 
This SOC enhancement is
also manifested in the increased magnitude of the effective $t_{\rm 2g}$ SOC
$\lambda_{t_{\text{2g}}}$ as shown in Table \ref{tab:hops}, which is 
obtained from the on-site matrix elements in the Wannier orbital calculations
\footnote{See \href{https://sites.google.com/site/heungsikim/home/temporary-files/SM_bLIO_resbmit_151023.pdf?attredirects=0&d=1}{Supplementary Materials} for further 
details.}.
Such behavior has also been reported in other 4$d$
and 5$d$ orbital systems \footnote{We note this SOC enhancement in
  \blio{} is less significant compared to other systems because the
  \jeff{} and $3/2$ states were already well-separated when
  $U_{\text{eff}}=0$~eV as a result of large
  SOC.}\cite{Liu2008SRO,hck_arxiv,Vijay2014}.

\begin{table}
  \centering
  \begin{tabular}{@{\extracolsep{\fill}}llrrr} \hline\hline 
    \multicolumn{2}{l}{$U_{\rm eff}$ (eV)}& ~~~~~~0.0 & ~~~~~~1.5 & ~~~~~~3.0 \\ \hline
    $\lambda_{t_{\rm 2g}}$      &&  0.401 &  0.482 &  0.516  \\
    $t_1$             & {\rm Z}  & +0.085 & +0.077 & +0.064 \\
                      & {\rm X}  & +0.083 & +0.074 & +0.058 \\
    $\vert t_2 \vert$ & {\rm Z}  &  0.238 &  0.255 &  0.270 \\
                      & {\rm X}  &  0.260 &  0.276 &  0.289 \\
    $t_3$             & {\rm Z}  & -0.162 & -0.119 & -0.060 \\
                      & {\rm X}  & -0.153 & -0.110 & -0.055 \\
    \hline\hline 
  \end{tabular}
  \caption{Magnitude of SOC within the Ir $t_{\rm 2g}$ states 
    and $t_{\rm 2g}$ hopping terms from Wannier orbital calculations
    in the presence of $U_{\rm eff}$. We adopt the coordinate system
    such that $t_2$ is negative for both Z and X bonds.  By symmetry,
    $t_2$ is positive for the X' bonds. 
  }
  \label{tab:hops}
\end{table}

The effects caused by electron correlations in the low-energy
\jeff{}-dominated states deserve further comment.
Fig. \ref{fig:jbands}(c) shows the schematic shape of Wannier orbitals
constructed from the \jeff{} energy window, which can be considered as
the local orbitals that span the low-energy subspace.  Owing to the
nearly ideal IrO$_6$ octahedra (as supported by the small amount of
trigonal distortion of less than 100~meV), the Wannier orbitals
consist of pure \jeff{} character on the center Ir site, while it has
$j_{\rm eff}=3/2$ tails on the NN sites. Similar
features have been reported in \aaio{N} and
\aaio{L}\cite{hsk2013LIO,Sohn2013mb}, which mirrors the remnant
molecular orbital character originating from the $t_{\rm 2g}$
hopping\cite{mazin_prl}.  As $U_{\rm eff}$ is included and
$\lambda_{t_{\rm 2g}}$ is enhanced, the \jeff{} character becomes more
dominant while $j_{\rm eff}=3/2$ components on the NN sites decreases
as shown in Fig. \ref{fig:jbands}(c) and (d).  The \jeff{}-like
Wannier orbital is more localized accordingly, which makes the
low-energy description of \blio{} in terms of the localized \jeff{}
states more feasible in the strong coupling limit.

\textit{$t_{\text{2g}}$ Wannier orbital hopping amplitudes --}
For a
detailed understanding of how the near-ideal structure of \blio{} is
manifested in the electronic band structure, we calculated the Ir
$t_{\text{2g}}$ hopping amplitudes from the Wannier orbitals in the
experimental structure. Table \ref{tab:hops} shows the magnitude of
the three largest hopping terms---$t_1$, $t_2$, and $t_3$---as the
value of $U_{\rm eff}$ changes ($U_{\text{eff}}=0.0$~eV, $1.5$~eV, and
$3.0$~eV, and SOC is included in the calculation); see
Sec. B and Fig. S1(a) in \href{https://sites.google.com/site/heungsikim/home/temporary-files/SM_bLIO_resbmit_151023.pdf?attredirects=0&d=1}{Supplementary Materials} and 
Ref. \cite{chkim_prl} for illustration of
these hopping processes. Since the Ir-Ir bond lengths and Ir-O-Ir bond
angles are similar on the two inequivalent bonds of \blio{} (X and Z
bonds), the values of their respective hopping amplitudes are expected
to be similar.  Indeed, by comparing the hopping amplitudes between
the two inequivalent NN bonds, we observe small anisotropies between
the X and Z bonds ($<10\%$),
which reflects the close-to-ideal structure of \blio{}.

The evolution of the NN hopping amplitudes as we include on-site Coulomb
interactions can be seen in Table \ref{tab:hops}. As $U_{\text{eff}}$
increases, $\vert t_2 \vert$ increases while $t_1$ and $t_3$ decrease.
Such behavior is understood in terms of the enhanced hybridization 
between the Ir $t_{\rm 2g}$ and oxygen $p$ states in the presence of
$U_{\rm eff}$. Inclusion of $U_{\text{eff}}$ pushes the
$j_{\text{eff}}$=3/2 states down energetically so that they become
closer to the oxygen $p$ states.
This leads to increased hybridization between the Ir $t_{\text{2g}}$
and oxygen $p$ states, which yields the enhancement of oxygen-mediated
$t_2$ (and the reduction of $t_1$ and $t_3$). 

\textit{Strong-coupling minimal model and experimental spiral phase --}
Having validated the use of the \jeff{} basis and the similarity of
hopping amplitudes between inequivalent bonds, we can now construct an
effective model to describe the low-energy properties of \blio{} in
the large-$U$ limit.  Following the derivation in
Ref. \cite{rau2014generic}, we start with localized \jeff{} states
then perform a strong-coupling expansion using NN $t_{\text{2g}}$
hopping amplitudes\footnote{Such {\it ab-initio} based perturbative approach 
applied to Na$_2$IrO$_3$\cite{Yamaji} yielded consistent results with the 
state-of-the-art quantum chemistry calculation performed in Ref. \cite{Katukuri}.}.  
In the presence of Hund's coupling $J_H$, we
arrive at a NN, \jeff{} model with highly anisotropic pseudospin
exchanges
\begin{equation*}
  H=\!\!\!\!\!\!\sum_{\langle i j \rangle \in  \alpha(\beta\gamma)} \!\!\!\! J^\alpha S_i \cdot
  S_j+K^\alpha S^\alpha_i S^\alpha_j+ \Gamma^\alpha
  (S^\beta_iS^\gamma_j + S^\gamma_iS^\beta_j),
\end{equation*}
where $S_i$ is the \jeff{} pseudospin on site $i$, $\alpha$ labels the
NN $\langle i j \rangle$ bond by its Kitaev component, and $\beta$ and
$\gamma$ denote the two non-Kitaev components of the $\langle i j
\rangle$-bond.  The exchanges $J$, $K$, and $\Gamma$ are functions of
the hopping amplitudes $t_1$-$t_3$, strength of Hund's coupling $J_H$,
SOC $\lambda$, and the on-site Coulomb interaction $U$: the relation
between these quantities are given in Sec. D in \href{https://sites.google.com/site/heungsikim/home/temporary-files/SM_bLIO_resbmit_151023.pdf?attredirects=0&d=1}{Supplementary Materials}.

To establish the region in the parameter space that best models
\blio{}, the following statistical analysis was employed. First, the 
hopping amplitudes and SOC values in Table \ref{tab:hops} 
were interpolated as a function of $U_{\text{eff}}$.  
Next, $U_{\rm eff}$ and $J_{\rm H}$ were treated as independent variables
with choices of ranges $1.5 < U_{\rm eff} < 3.0$ eV and
$\vert t_2 \vert < J_{\rm H} < \lambda_{t_{\rm 2g}}$ 
(note that $t_2$ is the largest hopping term)\footnote{The ranges of 
  $U_{\text{eff}}$ and $J_{\rm H}$ we used in this work are consistent with
  recent constrained RPA calculations of both parameters
  in the Ir 5$d$ orbitals\cite{cRPA,Yamaji}.},
due to the difficulty in
determining specific values of $U_{\rm eff}$ and $J_{\rm H}$. 
In order to present the phase evolution as a function of $U_{\rm eff}$, 
we chose six $U_{\rm eff}$ invervals centered at 1.5, 1.8, 2.1, 2.4, 2.7, and 3.0 eV with
ranges $\Delta U_{\rm eff}=\pm 0.1$ eV. From these parameters, six possible ranges
of the exchange parameters were estimated as shown in the inset of
Fig. \ref{fig:magldau}\footnote{For
  $U_{\text{eff}}$ and $J_{\rm H}$, we chose the triangular and uniform 
  distributions respectively for the stated ranges.  The qualitative features of the resulting
  exchange parameters' density distribution are not dependent on the
  precise distribution used but only on the mean and range of the
  distribution.}. 
We found that the mean anisotropies between X and Z bonds in $J$, $K$, and
$\Gamma$ are $3\%$, $15\%$, and $<1\%$ respectively relative to the
largest energy scale, which is the Kitaev exchange.  As a first
approximation, we treated all exchanges as isotropic between the X and Z
bonds, which yields the NN Hamiltonian studied in the classical limit
in Ref. \cite{Eric2014tm}. Lastly, we overlaied the density distributions
of the exchange parameters on top of the relevant
portion of the classical phase diagram reproduced from
Ref. \cite{Eric2014tm}, thereby yielding Fig. \ref{fig:magldau}. 

% The intensity of the overlay illustrates the frequency of appearance of those
% exchange parameters: the darker the overlay in a region is, the more frequently
% the parameter at the region appear in our estimation.
% Note that, the purpose of our approach is to set a reasonable
% range of exchange parameters to which the real material belongs:
% higher density regions are not necessarily more realistic.

% It does not necessarily mean that the values of exchange parameters 
% in the high-density regime are more realistic.

% Lastly, we overlay the joint probability
% density function of the exchange parameters on top of the relevant
% portion of the classical phase diagram reproduced from
% Ref. \cite{Eric2014tm}, thereby yielding Fig. \ref{fig:magldau}.  The
% intensity of the overlay illustrates the probability density: the
% darker the overlay in a region is, the higher its probability density.
% We interpret this probability density as the likelihood that the real
% material belongs in that region of parameter space.

The phase diagram is the quarter of a polar plot near the ferromagnetic
Kitaev limit: The angular
coordinate $\phi$ shows the ratio between $J$ and
$K$ --- $\tan(\phi)=J/K$.  Meanwhile, the radial coordinate depicts the
strength of $\Gamma$ --- $\theta \in(\pi/2, \pi)$,
$\tan\theta=\sqrt{J^2+K^2}/\Gamma$.  The bottom boundary of the
half-ring ($\theta=\pi/2$, $\phi = 3\pi/2$) is the Heisenberg-Kitaev 
limit ($\Gamma=0$) and the origin ($\theta=\pi$)
is the pure $\Gamma$ limit.  As seen in the figure, the NN exchanges
in \blio{} are likely dominated by a large, ferromagnetic Kitaev
exchange, perturbed by small $|J|$ and $|\Gamma|$.
Note that, higher $U_{\rm eff}$ and $J_{\rm H}$ prefer smaller $\theta$ 
(larger $K$) and smaller $\phi$, respectively.

We find that, when $U_{\rm eff} \geq 2.4$ eV, the correspondins phases lie within 
area of the spiral phase $\overline{\text{SP}}_{a^{\textrm{-}}}$.
Remarkably, this non-coplanar spiral magnetic phase
possesses the same symmetries as the experimentally determined
magnetic order.\cite{Eric2014tm,Biffin2014ky} In other words, using
the \textit{ab-initio} hopping and SOC parameters, the resulting
exchange parameters in the isotropic $J$-$K$-$\Gamma$ pseudospin model
results in a classical ground state that agrees with the experimental
magnetic order.

\textit{Discussion and Conclusion --}
Although we have shown that the
\jeff{} states form a valid basis as a consequence of the small amount
of distortions present, the difference between the dispersions of the
ideal and experimental structures away from the Fermi level are due to
these distortions and the resulting bond anisotropies.  However, in
the context of the effective pseudospin model, we have shown that
these non-idealities are negligible for the $J$ and $\Gamma$
exchanges.  The Kitaev exchange, on the other hand, is more
anisotropic between the X and Z bonds, but we speculate that the
$\overline{\text{SP}}_{a^-}$ spiral phase will remain robust under
this anisotropy; we leave the investigation on the effects of bond
anisotropy for future work.  Nevertheless, it remains true that the NN
exchanges on both X and Z bonds are dominated by large ferromagnetic
Kitaev exchanges and that the Kitaev spin liquid is robust against
bond-anisotropies.\cite{kitaev2006anyons}

In addition to distortions and bond anisotropies, an accurate
description of the electronic structure also requires hopping
amplitudes beyond the NN level
(see Sec. C in \href{https://sites.google.com/site/heungsikim/home/temporary-files/SM_bLIO_resbmit_151023.pdf?attredirects=0&d=1}{Supplementary Materials} for details).
These terms would generate further neighbor exchange
interactions in the strong coupling theory. 
% Additionally, 
% higher-order superexchange processes can be possible, which will
% lead to further neighbor interactions as well.
However, these exchanges are no more than $10\%$ of those at the NN level. 
Since we expect that such small further-neighbor interactions
do not change our conclusions, we focused on the NN exchange interactions
in out manuscript.

To enhance the Kitaev exchange relative to other interactions and to
approach the spin-liquid regime of the Kitaev model, strengthening the
oxygen-mediated-type hopping ($t_2$) is a viable option.  Increasing
the on-site Coulomb interaction can further localize the
$t_{2\text{g}}$ orbitals, which reduces the amplitudes for direct
hopping channels like $t_1$ and $t_3$ while oxygen-mediated hopping
channels like $t_2$ are comparatively less affected.  In addition,
increasing $U$ has the effect of driving the system deeper into the
Mott insulating regime and reducing the strength of further neighbor
interactions.  Therefore, a $4d$ variant of the \blio{} may offer the
right ingredients to enhance the Kitaev exchange.

Indeed, the isoelectronic, 2D honeycomb $\alpha$-Li$_2$RhO$_3$ has
been synthesized and argued to be a relativistic Mott insulator
driven by electronic correlations and SOC.\cite{luo2013li}
Furthermore, this material does not magnetically order down to 0.5~K,
which is an indication of magnetic frustration.\cite{luo2013li} We
speculate that the hypothetical 3D polymorph---hyperhoneycomb
$\beta$-Li$_2$RhO$_3$---may be a less distorted version of
$\alpha$-Li$_2$RhO$_3$ that has all the right properties to further
approach the Kitaev region, in analogy to \blio{}.

\textit{Acknowledgments --}
This work was supported by the NSERC of
Canada and the center for Quantum Materials at the University of
Toronto.  Computations were mainly performed on the GPC supercomputer
at the SciNet HPC Consortium. SciNet is funded by: the Canada
Foundation for Innovation under the auspices of Compute Canada; the
Government of Ontario; Ontario Research Fund - Research Excellence;
and the University of Toronto.  HSK thanks to IBS Center for
Correlated Electron System in Seoul National University for additional
computational resources.

%\bibliography{bLIO}
% section acknowledgements (end)

\appendix

\setcounter{figure}{0}
\makeatletter 
\renewcommand{\thefigure}{S\@arabic\c@figure}
\makeatother

\setcounter{table}{0}
\makeatletter 
\renewcommand{\thetable}{S\@arabic\c@figure}
\makeatother

\begin{figure*}[htb!]
  \centering
  \includegraphics[width=0.9 \textwidth]{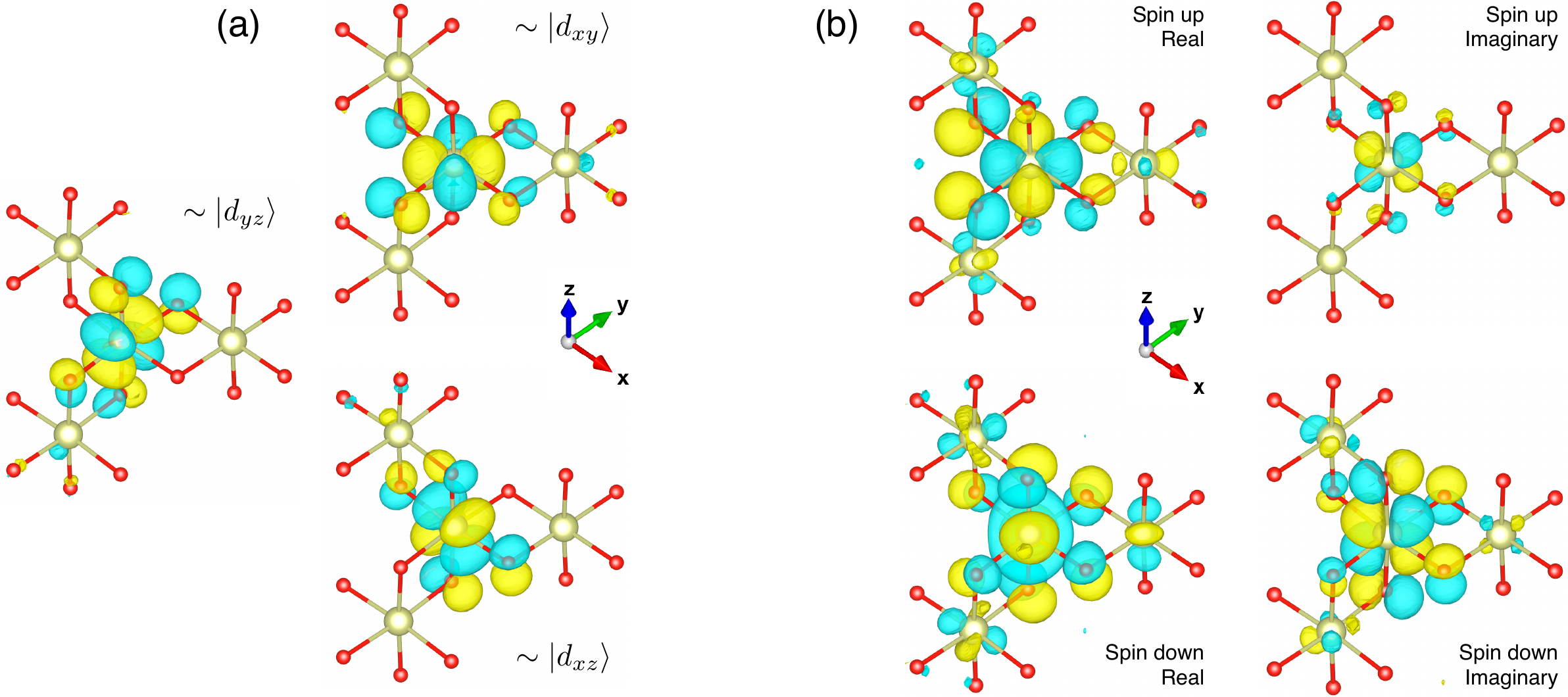}
  \caption{(Color online) (a) Ir $t_{\rm 2g}$ Wannier orbitals for up spin component 
  located at an Ir site, obtained in a GGA+SOC calculation. Other 21 $t_{\rm 2g}$
  Wannier orbitals can be obtained by translating them to other Ir sites in
  the unit cell and applying time-reversal operation. Only nearest-neighboring Ir atoms are shown. 
  (b) One Ir $j_{\rm eff}$=1/2 Wannier orbital obtained in a PBE+SO+$U$ 
  calculation with $U_{\rm eff}$ = 3 eV. Other eight $j_{\rm eff}$=1/2
  Wannier orbitals can be obtained by translations and time-reversal operation.
  Note that, isosurface value for drawing the $t_{\rm 2g}$ Wannier orbital in (a) 
  is twice larger than the one used in (b). 
  }
  \label{figA:Wannier}
\end{figure*}

\section{\label{app:dft_details}Supplementary Material A: \\
Details on \textit{ab-initio} electronic structure
  calculations}
For the electronic structure calculations with SOC and on-site Coulomb
interaction, OPENMX code\cite{openmx,*openmx2}, which is based on the
linear-combination-of-pseudo-atomic-orbital basis formalism, was
used. A non-collinear DFT scheme and a fully relativistic
$j$-dependent pseudopotential were used to treat SOC, and
Perdew-Burke-Ernzerhof parametrization of the generalized
gradient approximation (GGA) was chosen for the exchange-correlation
functional\cite{PBE}, which was compared and found to be
almost identical with the results with the Perdew and Zunger local
density approximation functional\cite{CA,*PZ}. 400 Ry of energy
cutoff was used for the real-space sampling, and $9\times9\times9$
$k$-grid was adopted for the primitive unit cell. Electron
interactions are treated as on-site Coulomb interactions via a
simplified LDA+$U$ formalism implemented in OPENMX
code\cite{han2006n}, and up to 3.0~eV of 
$U_{\rm eff} \equiv U-J_{\rm H}$ parameter ($J_{\rm H}$ is 
Hund's coupling) was used for Ir $d$
orbital in our GGA+SOC+$U$ calculations.  Maximally-localized Wannier
orbital method\cite{marzari1997maximally}, which is implemented in
OPENMX code\cite{weng2009revisiting}, were used to obtain the
tight-binding Hamiltonian for Ir $t_{\rm 2g}$ atoms.

\section{\label{app:wann}Supplementary Material B: \\
Ir $t_{\rm 2g}$ and $j_{\rm eff}$=1/2 Wannier orbitals}
In order to obtain the hopping integrals between the Ir $t_{\rm 2g}$ states, 
the Wannier orbitals for the Ir $t_{\rm 2g}$ bands were calculated in the presence 
of SOC. Fig. \ref{figA:Wannier} shows the results, where the three $t_{\rm 2g}$ 
Wannier orbitals at an Ir site with spin up component are shown. Other 21 orbitals 
at the four Ir sites in the unit cell are obtained by translating them and applying
time-reversal operation $i\sigma_y \mathcal{K}$, where $\sigma_y$ and $\mathcal{K}$ are
the Pauli matrix acting on the spin sector and the complex conjugation, respectively.
The oxygen $p$-orbital components hybridized into the Ir
$t_{\rm 2g}$ bands are manifested as the oxygen $p$-orbital tails shown in the figure. 
Size of oxygen hybridization is slightly enhanced as the value of $U_{\rm eff}$ is increased,
which contribute to the $t_2$ hopping term dominated by the oxygen-mediated channel. 

The values of the effective SOC strength $\lambda_{t_{\rm 2g}}$ mentioned in the main text 
were estimated from the on-site energies between the $t_{\rm 2g}$ Wannier orbitals. 
The on-site energy matrix can be approximately expressed as 
$H_{\rm on} \approx \lambda_{t_{\rm 2g}} \boldsymbol{l}_{t_{\rm 2g}} \cdot \boldsymbol{s} + \Delta_t$, 
where $\Delta_t$ is the minor trigonal crystal field terms. The value of $\lambda_{t_{\rm 2g}}$ 
for each $U_{\rm eff}$ value 
was obtained by taking the average of the matrix elements corresponding to the SOC term, 
with their standard deviation smaller than 10\% of their average.

Fig. \ref{figA:Wannier}(b) shows the $j_{\rm eff}$=1/2-like Wannier orbital, 
obtained from the low-energy window dominated by the $j_{\rm eff}$=1/2 character as 
illustrated in Fig. 3(c) in the main text,
in a GGA+SOC+$U$ calculation with $U_{\rm eff}$ = 3 eV. Like the $t_{\rm 2g}$ orbitals, 
other seven $j_{\rm eff}$=1/2-like Wannier orbitals in the unit cell can be obtained by 
translation and time reversal operations.
Decomposing the Wannier orbital in terms of the local Ir $t_{\rm 2g}$ basis reveals 
the dominant $j_{\rm eff}$=1/2 character at the center with
the $j_{\rm eff}$=3/2-dominated tails on the three next-neighboring Ir sites, as 
schematically shown in Fig. 3(c) in the main text. 
The tail components are gradually reduced as the $U_{\rm eff}$ value is increased, 
so that the Wannier orbital becomes more localized in the presence of higher $U_{\rm eff}$.

\begin{figure*}[htb!]
  \centering
  \includegraphics[width=0.9 \textwidth]{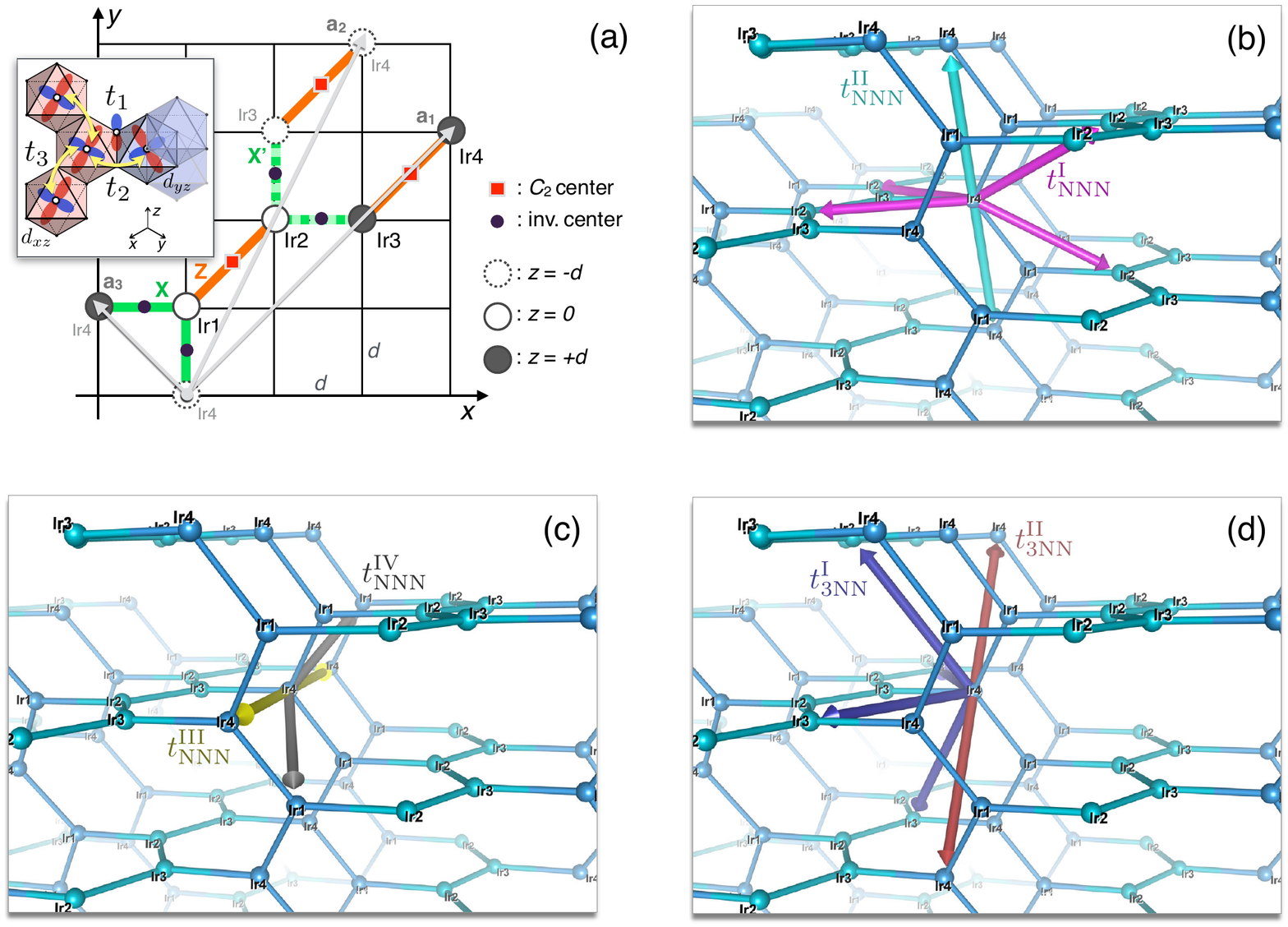}
  \caption{(Color online) (a) Ir sites projected onto $xy$-plane
  defined in terms of local cubic axes. ${\bf a}_1$, ${\bf a}_2$
  and ${\bf a}_3$ are the Bravais lattice vector for the primitive cell,
  where four sublattices within the primitive cell are labeled as Ir1 to Ir4. 
  Centers for $C^{a,b,z}_2$ rotations and inversion are depicted as 
  red square and black dots on the Z and X (X') bonds, respectively.
  Note that $d$ becomes same with the Ir-O bond length in the absence of 
  trigonal distortion. Inset shows the three major hopping channels 
  between NN Ir sites.
  (b,c) All NNN and (d) third NN neighbors
  for an Ir site, Ir4, are depicted as colored arrows. 
  Note that different colors (blue and green) are used to distinguish 
  Ir sites and bonds belonging to one zigzag chain to another.
  (a) shows NNN hopping paths that can be reached through intermediate NN
  Ir sites. Intra- and inter-chain bonds are colored as cyan and purple,
  respectively. 
  (b) shows NNN paths that cannot be reached through one NN bonds.
  Yellow and grey arrows represent paths connecting same (Ir4 to Ir4)
  and different sublattices (Ir4 to Ir1 in the figure), respectively.
  (c) shows all third NN hopping paths. Purple and red arrows show bonds 
  connecting different and same sublattices, respectively.
  Note that, bonds in (b) and red bonds in (c) does not have 
  any counterparts in the 2D honeycomb lattice.
  }
  \label{figA:NNNN}
\end{figure*}

\section{\label{app:hops}Supplementary Material C: \\
$t_{\rm 2g}$ hopping terms and tight-binding bands from Wannier orbitals}

Table \ref{tabA:hops} shows a partial list of Ir $t_{\rm 2g}$ hopping terms (up to third NN)
from the Wannier orbitals, where the convention for the coordinate system and the 
illustration of NNN and third NN hopping terms are in Fig. \ref{figA:NNNN}. 
Full list of hopping terms can be restored by applying the $C_2$ rotations and 
inversion operations at the centers of Z and X (X') bonds, respectively. 
Three $C_2$ rotations --- $C^{a,b,z}_2$ --- are allowed,
where $a \equiv \hat{x} + \hat{y}$ and $b \equiv \hat{y} - \hat{x}$. 

Contrary to the relatively simple NN hopping channels as shown in the inset of 
Fig. \ref{figA:NNNN}(a), a number of distinct NNN hopping
terms are introduced due to the three-dimensional twisting of the honeycomb
lattice\cite{Eric2014band}.
The NNN hopping channels can be classified into two kinds, depending on whether
they are analogous to the NNN hopping in the 2D honeycomb lattice or not.
Fig. \ref{figA:NNNN}(b) shows the 2D-like NNN hoping channels, which can be
reached through one intermediate NN Ir site. Depending on whether they belong to same zigzag
chain composed of only X (or X') bonds or connect different chains, they are divided into two different classes 
$t^{\rm I}_{\rm NNN}$ and $t^{\rm II}_{\rm NNN}$. Hopping amplitudes in these channels are 
larger than the other channels, $t^{\rm III}_{\rm NNN}$ and $t^{\rm IV}_{\rm NNN}$, 
which cannot be reached through one Ir site as shown in Fig. \ref{figA:NNNN}(c). 
There are also non-negligible third NN hopping terms, $t^{\rm I}_{\rm 3NN}$ 
and $t^{\rm II}_{\rm 3NN}$, which can be seen in Fig. \ref{figA:NNNN}(d) and 
Table \ref{tabA:hops}. Like NNN hopping channels, third NN channels can be classified
depending on whether they have their 2D counterparts or not. $t^{\rm I}_{\rm 3NN}$ 
resembles the third NN hopping channel in the 2D honeycomb lattice, while 
$t^{\rm II}_{\rm 3NN}$ is similar to the interlayer hopping terms in \aaio{$A$} series. 

\begin{figure*}[htb!]
  \centering
  \includegraphics[width=0.7\textwidth]{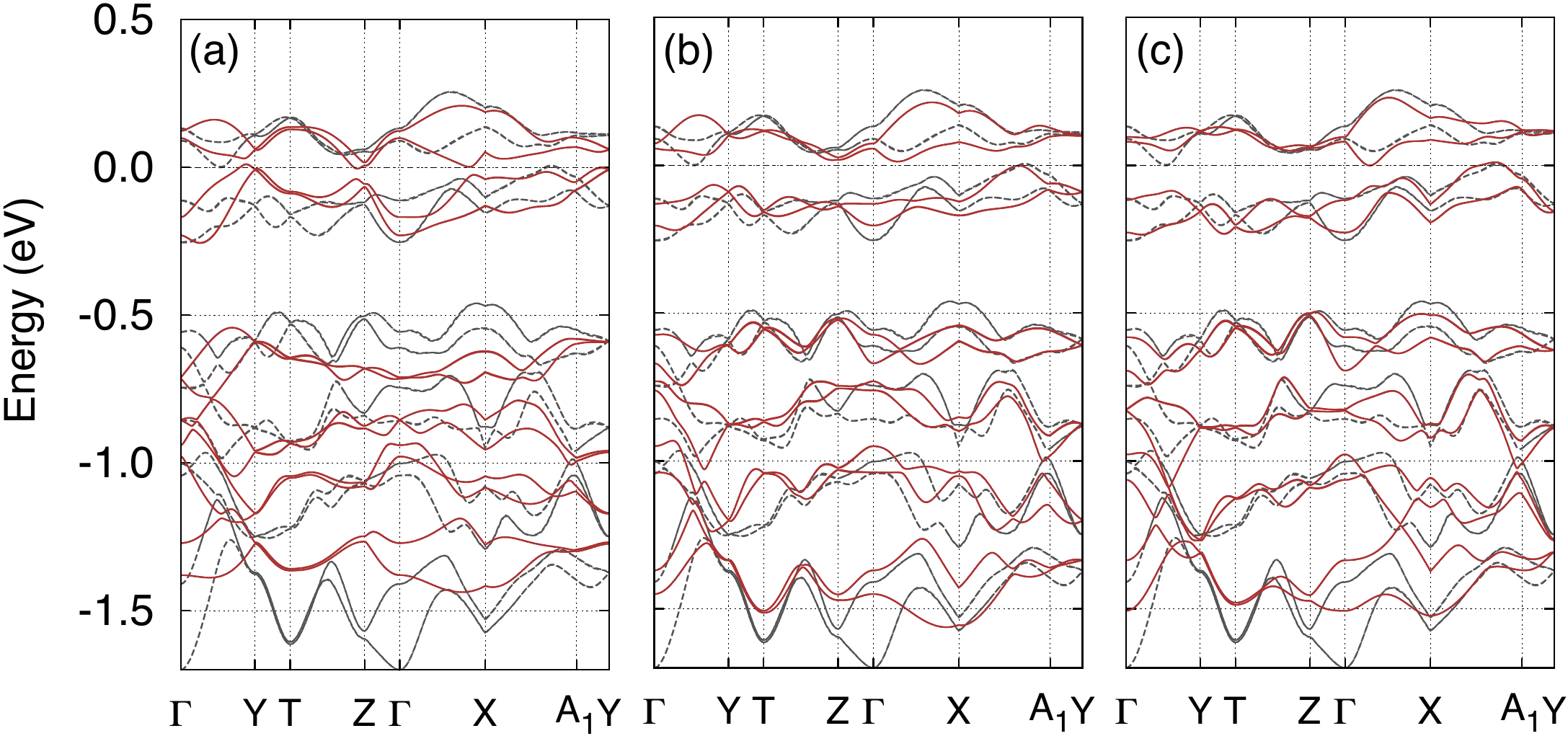}
  \caption{(Color online) 
  Band structures (solid red lines) calculated from tight-binding calculations including up to
  (a) NN, (b) NNN, and (c) third-NN hopping terms from the Wannier orbital calcualtions
  tabulated in Table. \ref{tabA:hops}, compared to those from {\it ab-initio} results
  (dashed grey lines). SOC is included in the calculations.
  }
  \label{figA:bands}
\end{figure*}

The role of further-neighbor hopping amplitudes in the band structure are shown in Fig. \ref{figA:bands}, 
where the evolution of band structure as we include NNN and third NN terms are presented. 
Fig. \ref{figA:bands} shows the change of the band structures as further-neighbor
hopping terms are included with the presence of SOC; Fig. \ref{figA:bands}(a), (b) and (c)
show the bands with hopping terms up to NN, NNN, and third NN terms, respectively, with the
presence of SOC. 
One can see that, the large SOC in Ir tends to make the \jeff{} bands to be flatter in this 
locally honeycomb-like lattice, and including NNN and third NN terms does not change the overall
behavior. Comparing Fig. \ref{figA:bands}(a) and Fig. \ref{figA:bands}(b), 
one can notice that the dispersion inside the \jeff{} subbands is affected by the NNN terms, 
but the semi-metallic character is left unchanged. Inclusion of third NN terms, 
as can be seen in Fig. \ref{figA:bands}(c), makes the dispersion slightly closer to the {\it ab-initio} bands.

\section{\label{app:exchanges}Supplementary Material D:\\
NN exchange interactions}
The exchanges $J$, $K$, and $\Gamma$ are given by (suppressing the
bond label $\alpha$)
\begin{align}
  J\!&=\! \frac{4}{27}\left[
    \frac{(2t_1+t_3)^2(4J_{H}+3U)}{U^2} - \frac{16J_{H}(t_1-t_3)^2}{(2U+3\lambda)^2}\right]\nonumber \\
  K\!&=\! \frac{32J_H}{9}\!\!\left[
    \frac{(t_1-t_3)^2\!-\!3t^2_2}{(2U+3\lambda)^2}\right]\!,\quad \!\!\!\!\!\! \Gamma= \frac{64J_H}{9}
    \frac{t_2(t_1-t_3)}{(2U+3\lambda)^2},
    \label{eq:exch}
\end{align}
where $t_i~(i=1,2,3)$, $J_H$, $U$, and $\lambda$ are the NN hopping
amplitudes, Hund's coupling, on-site Coulomb repulsion, and SOC
respectively\cite{rau2014generic}. $t_i$ is illustrated in Fig. \ref{figA:NNNN}(a). 
Note that, the small amount of NN Heisenberg interaction is attributed to the cancelation between the
$2t_1$ and $t_3$ in the antiferromagnetic contrubution to $J$ in Eq. 1. 
Since $t_2$ is the largest term, as mentioned in the main text, ferromagnetic $K$ becomes
the most dominant contribution in the exchange interactions.

\renewcommand*{\arraystretch}{1.2}
\begin{longtable*}{@{\extracolsep{\fill}}llllrrrrrrrrr} 
\caption{A subset of Ir $t_{\rm 2g}$ hopping terms ${\bf T}_{ij}$ as representatives of each hopping channels up to third NN,
where $\mathcal{H}_{\rm hop} = \sum_{ij} {\bf C}^{\dag}_{i} \cdot {\bf T}_{ij} \cdot {\bf C}_j$
and ${\bf C}^{\dag}$ and ${\bf C}^{\dag}$ being the creation and annihilation operator for $t_{\rm 2g}$ states, respectively.
$d$ is approximate distance between Ir and O. Other hopping terms can be recovered by applying ${\bf T}_{ji} = {\bf T}^{\dag}_{ij}$,
$C^{a,b,z}_{2}$ rotations, and inversion operations.} \\
\hline\hline 
Kind &${\bf r}_{ij}$ (in Cartesian coord.) & Sublattice & & \multicolumn{3}{r}{$U_{\rm eff}$ = 0.0~eV} & \multicolumn{3}{r}{$U_{\rm eff}$ = 1.5~eV} & \multicolumn{3}{r}{$U_{\rm eff}$ = 3.0~eV} \\ [5pt]
\hline
\endfirsthead
\multicolumn{13}{r}{{Continued from previous page...}} \\ \hline\hline
Kind &${\bf r}_{ij}$ (in Cartesian coord.) & Sublattice & & \multicolumn{3}{r}{$U_{\rm eff}$ = 0.0~eV} & \multicolumn{3}{r}{$U_{\rm eff}$ = 1.5~eV} & \multicolumn{3}{r}{$U_{\rm eff}$ = 3.0~eV} \\ [5pt] \hline
\endhead
\hline \multicolumn{13}{r}{{Continued in next page...}} \\ \hline
\endfoot
\hline\hline 
\endlastfoot

$t_{\rm NN}$ &&&& ~~~~~~~~~~ $d_{xy}$ & $d_{xz}$ & $d_{yz}$ & ~~~~~~~~~~ $d_{xy}$ & $d_{xz}$ & $d_{yz}$ & ~~~~~~~~~~ $d_{xy}$ & $d_{xz}$ & $d_{yz}$ \\
X,X'&(-$d$,~0,+$d$) & $1 \rightarrow 4$ & $d_{xy}$ &  +0.088 &  +0.018 &  +0.260 &  +0.080 &  +0.019 &  +0.276 &  +0.064 &  +0.021 &  +0.289 \\
&                                            &&$d_{xz}$&  +0.018 &  -0.152 &  +0.013 &  +0.020 &  -0.110 &  +0.013 &  +0.021 &  -0.051 &  ~0.005 \\
&                                            &&$d_{yz}$&  +0.259 &  +0.013 &  +0.078 &  +0.276 &  +0.013 &  +0.067 &  +0.288 &  ~0.003 &  +0.052 \\[5pt]
$t_{\rm NN}$ &&&& ~~~~~~~~~~ $d_{xy}$ & $d_{xz}$ & $d_{yz}$ & ~~~~~~~~~~ $d_{xy}$ & $d_{xz}$ & $d_{yz}$ & ~~~~~~~~~~ $d_{xy}$ & $d_{xz}$ & $d_{yz}$ \\
Z   &(+$d$,+$d$,~0) & $1 \rightarrow 2$ & $d_{xy}$ &  -0.162 &  -0.022 &  +0.021 &  -0.119 &  -0.024 &  +0.023 &  -0.059 &  -0.031 &  +0.030 \\
&                                            &&$d_{xz}$&  +0.016 &  +0.087 &  -0.239 &  +0.017 &  +0.078 &  -0.255 &  +0.025 &  +0.072 &  -0.269 \\
&                                            &&$d_{yz}$&  -0.016 &  -0.239 &  +0.086 &  -0.017 &  -0.254 &  +0.077 &  -0.024 &  -0.271 &  +0.056 \\[5pt]
$t^{\rm I}_{\rm NNN}$&& & & ~~~~~~~~~~ $d_{xy}$ & $d_{xz}$ & $d_{yz}$ & ~~~~~~~~~~ $d_{xy}$ & $d_{xz}$ & $d_{yz}$ & ~~~~~~~~~~ $d_{xy}$ & $d_{xz}$ & $d_{yz}$ \\ 
&(+$d$,+$2d$,-$d$) & $1 \rightarrow 3$ & $d_{xy}$ &  ~0.002 &  -0.012 &  +0.039 &  ~0.001 &  -0.015 &  +0.044 &  ~0.001 &  -0.018 &  +0.047 \\
&                                            &&$d_{xz}$&  +0.013 &  ~0.001 &  +0.011 &  +0.018 &  ~0.001 &  +0.014 &  +0.024 &  ~0.001 &  +0.017 \\
&                                            &&$d_{yz}$&  +0.063 &  ~0.004 &  ~0.002 &  +0.075 &  ~0.007 &  ~0.000 &  +0.089 &  -0.010 &  ~0.001 \\[5pt]
$t^{\rm II}_{\rm NNN}$&& & & ~~~~~~~~~~ $d_{xy}$ & $d_{xz}$ & $d_{yz}$ & ~~~~~~~~~~ $d_{xy}$ & $d_{xz}$ & $d_{yz}$ & ~~~~~~~~~~ $d_{xy}$ & $d_{xz}$ & $d_{yz}$ \\ 
&(-$d$,+$d$,+$2d$) & $1 \rightarrow 1$ & $d_{xy}$ &  ~0.002 &  ~0.008 &  -0.014 &  ~0.003 &  -0.011 &  -0.017 &  ~0.003 &  -0.014 &  -0.020 \\
&                                         &&$d_{xz}$&  +0.014 &  ~0.001 &  +0.039 &  +0.017 &  ~0.000 &  +0.045 &  +0.020 &  ~0.002 &  +0.050 \\
&                                         &&$d_{yz}$&  ~0.008 &  +0.075 &  ~0.001 &  +0.011 &  +0.089 &  ~0.000 &  +0.014 &  +0.103 &  ~0.001 \\[5pt]
$t^{\rm III}_{\rm NNN}$&& & & ~~~~~~~~~~ $d_{xy}$ & $d_{xz}$ & $d_{yz}$ & ~~~~~~~~~~ $d_{xy}$ & $d_{xz}$ & $d_{yz}$ & ~~~~~~~~~~ $d_{xy}$ & $d_{xz}$ & $d_{yz}$ \\ 
&(-$d$,+$d$,-$2d$) & $1 \rightarrow 1$ & $d_{xy}$ &  ~0.001 &  +0.011 &  ~0.001 &  ~0.000 &  +0.013 &  ~0.000 &  ~0.002 &  +0.015 &  ~0.003 \\
&                                            &&$d_{xz}$&  ~0.001 &  ~0.007 &  +0.038 &  ~0.000 &  ~0.005 &  +0.045 &  ~0.003 &  ~0.002 &  +0.051 \\
&                                            &&$d_{yz}$&  -0.011 &  +0.030 &  ~0.008 &  -0.013 &  +0.036 &  ~0.006 &  -0.015 &  +0.047 &  ~0.004 \\[5pt]
$t^{\rm IV}_{\rm NNN}$&& & & ~~~~~~~~~~ $d_{xy}$ & $d_{xz}$ & $d_{yz}$ & ~~~~~~~~~~ $d_{xy}$ & $d_{xz}$ & $d_{yz}$ & ~~~~~~~~~~ $d_{xy}$ & $d_{xz}$ & $d_{yz}$ \\ 
&(+$d$,-$2d$,+$d$) & $1 \rightarrow 4$ & $d_{xy}$ &  +0.012 &  ~0.007 &  -0.030 &  ~0.009 &  ~0.008 &  -0.035 &  ~0.005 &  ~0.008 &  -0.041 \\
&                                            &&$d_{xz}$&  ~0.007 &  ~0.008 &  ~0.006 &  ~0.008 &  ~0.009 &  ~0.008 &  ~0.008 &  +0.011 &  ~0.008 \\
&                                            &&$d_{yz}$&  -0.030 &  ~0.007 &  ~0.003 &  -0.035 &  ~0.008 &  ~0.007 &  -0.041 &  ~0.008 &  -0.012 \\[5pt]
$t^{\rm I}_{\rm 3NN}$&& & & ~~~~~~~~~~ $d_{xy}$ & $d_{xz}$ & $d_{yz}$ & ~~~~~~~~~~ $d_{xy}$ & $d_{xz}$ & $d_{yz}$ & ~~~~~~~~~~ $d_{xy}$ & $d_{xz}$ & $d_{yz}$ \\ 
&(~0,+$2d$,-$2d$) & $1 \rightarrow 2$ & $d_{xy}$ &  ~0.007 &  -0.013 &  -0.014 &  ~0.007 &  -0.014 &  -0.015 &  ~0.008 &  -0.015 &  -0.017 \\
&                                            &&$d_{xz}$&  ~0.008 &  ~0.006 &  ~0.006 &  ~0.009 &  ~0.008 &  ~0.007 &  -0.011 &  +0.011 &  ~0.009 \\
&                                            &&$d_{yz}$&  -0.014 &  -0.016 &  -0.034 &  -0.015 &  -0.018 &  -0.030 &  -0.017 &  -0.020 &  -0.022 \\[5pt]
$t^{\rm II}_{\rm 3NN}$&& & & ~~~~~~~~~~ $d_{xy}$ & $d_{xz}$ & $d_{yz}$ & ~~~~~~~~~~ $d_{xy}$ & $d_{xz}$ & $d_{yz}$ & ~~~~~~~~~~ $d_{xy}$ & $d_{xz}$ & $d_{yz}$ \\ 
&(-$d$,+$d$,~0) & $1 \rightarrow 1$ & $d_{xy}$ &  -0.045 &  ~0.006 &  -0.012 &  -0.046 &  ~0.007 &  -0.013 &  -0.046 &  -0.010 &  -0.013 \\
&                                            &&$d_{xz}$&  +0.012 &  ~0.008 &  ~0.001 &  +0.013 &  ~0.009 &  ~0.001 &  +0.013 &  +0.012 &  ~0.002 \\
&                                            &&$d_{yz}$&  ~0.006 &  -0.015 &  ~0.008 &  ~0.008 &  -0.017 &  ~0.009 &  +0.010 &  -0.018 &  +0.012
\label{tabA:hops}
\end{longtable*}

\bibliography{bLIO}

\end{document}